\pdfoutput=1
\documentclass{raa_twocolumn}
\usepackage{graphicx,times}
\usepackage{natbib}
\usepackage{amssymb,amsmath}
\usepackage[OT1]{fontenc}
\usepackage{tabularx}
\usepackage{subfigure}
\usepackage{url}
\usepackage{xtab}
\usepackage{longtable}
\usepackage{threeparttable}
\usepackage[colorlinks,
linkcolor=blue,
anchorcolor=blue,
citecolor=blue,
]{hyperref}

\bibpunct{(}{)}{;}{a}{}{,}

\usepackage{tabularx} 
\usepackage{ragged2e} 
\usepackage{booktabs} 

\newcommand{\Eq}{{Equation}}

\newcommand{\Fig}{{Figure}}

\begin{document}

\title{MHD simulations of the slow-rise phase of solar eruptions initiated from a sheared magnetic arcade}

\volnopage{ {\bf 20XX} Vol.\ {\bf X} No. {\bf XX}, 000--000} \setcounter{page}{1}

\author{Qingjun Liu, Chaowei Jiang\inst{*}, Zhipeng Liu}

\institute{Shenzhen Key Laboratory of Numerical Prediction for Space Storm, School of Aerospace Science,\\ Harbin Institute of Technology, Shenzhen 518055, China; \\
*chaowei@hit.edu.cn}

\abstract{
Before solar eruptions, a short-term slow-rise phase is often observed, during which the pre-eruption structure ascends at speeds much greater than the photospheric motions but much less than those of the eruption phase. Numerical magnetohydrodynamic (MHD) simulations of the coronal evolution driven by photospheric motions up to eruptions have been used to explain the slow-rise phase, but their bottom driving speeds are much larger than realistic photospheric values. Therefore, it remains an open question how the excessively fast bottom driving impacts the slow-rise phase. Here we modelled the slow-rise phase before eruption initiated from a continuously sheared magnetic arcade. In particular, we performed a series of experiments with the bottom driving speed unprecedentedly approaching the photospheric value of around $1$~km~s$^{-1}$. The simulations confirmed that the slow-rise phase is an ideal MHD process, i.e., a manifestation of the growing expansion of the sheared arcade in the process of approaching a fully open field state. The overlying field line above the core flux has a slow-rise speed modulated by the driving speed's magnitude but is always over an order of magnitude larger than the driving speed. The core field also expands with speed much higher than the driving speed but much lower than that of the overlying field. By incrementally reducing the bottom-driving speed to realistic photospheric values, we anticipate better matches between the simulated slow-rise speeds and some observed ones.
}

\authorrunning{Liu et al.}            
\titlerunning{Simulation of slow-rise phase}  
\maketitle

\section{Introduction}
\label{sect:intro}

Solar eruptions, encompassing phenomena such as solar flares and coronal mass ejections (CMEs), are among the most dynamic and powerful events in the solar system~\citep{2003NewAR..47...53L, 2003A&A...397.1057Z, 2006A&A...445.1133Z, chenCoronalMassEjections2011, 2024ScChD..67.3765J}. These events are characterized by the sudden release of vast amounts of magnetic free energy stored in the coronal magnetic field, resulting in dramatic changes to solar and interplanetary environments. Understanding the initiation mechanisms behind these eruptions is not only fundamental to solar physics~\citep{2017ScChD..60.1383C, 2017ScChD..60.1408G,2019ApJ...884L..51Y,2021NatCo..12.2734Z,2022NatCo..13..640Y} but also critical for predicting space weather events that can have profound impacts on the Earth's technological infrastructures~\citep{2024NatCo..15.9198T}.

The initiation of CMEs is often associated with a slow-rise phase~\citep{1988ApJ...328..824K, Joshi2017, 2020ApJ...894...85C}, which occurs after the quasi-static evolving state of CME progenitors but right before the impulsively eruptive state. This phase is closely linked to the eruption mechanism, although its physical nature remains unclear. The slow-rise phase is characterized by coronal magnetic field expansion, which is indicated by motion of filaments~\citep[e.g.,][]{2010SSRv..151..333M,Guo2019}, hot channels~\citep[e.g.,][]{Cheng2013,Yao2024} and overlying loops~\citep[e.g.,][]{10.1093/mnras/stae2088}, with a typical speed (of a few tens~km~s$^{-1}$) much larger than that of the photospheric motion ($\leqslant 1$~km~s$^{-1}$) but much smaller than the eruption speed (hundreds to a few thousands km~s$^{-1}$). Therefore, it represents a unique intermediate state that is neither fully quasi-static nor eruptive.

Currently there exists two distinct explanations for the slow-rise phase, which may fit different circumstances depending on the specific magnetic configurations. Based on some observations~\citep{Cheng2023} and an MHD simulation~\citep{Aulanier2010} of the formation and eruption of a magnetic flux rope (MFR), \citet{Xing2024} suggested that the slow-rise phase results from slow tether-cutting magnetic reconnection below the MFR shortly before eruption. This reconnection not only weakens the overlying field but also drives the slow ascent of the newly formed MFR due to the upward magnetic tension from the reconnected field lines. Since the reconnection is slow, the energy release is gradual, leading to a slow rise of the MFR compared to the eruption phase. In the case of an eruption initiated from sheared arcade (rather than gradually formed pre-eruption MFR), \cite{10.1093/mnras/stae2088} suggested that the slow-rise phase is an purely ideal MHD process (i.e., not associated with reconnection) based on the simulation by~\cite{2021NatAs...5.1126J} in which reconnection within a sheared magnetic arcade leads to eruption. They suggested that the slow-rise phase is a manifestation of the growing expansion of the arcade in the process of approaching a fully open field state, which is inherent to the formation of a current sheet before the eruption. In this scenario, reconnection begins only at the onset of the eruption, and in particular, the slow rise is more prominently demonstrated by the overlying field, which has a rising speed significantly larger than that of the core field. \citet{10.1093/mnras/stae2088} further investigated some flare events with slow-rise phase and found that they are consistent with the simulation: the overlying coronal loops presents an expansion much faster than the slow rise of filament that represents the core field.

Nevertheless, in the aforementioned MHD simulations~\cite[and essentially all the currently available MHD simulations that are designed to mimic the coronal magnetic field energization driven by photospheric motions until eruptions, e.g., ][]{Amari2003,DeVore2008,Aulanier2010}, the driving speed applied at the bottom boundary, e.g., the often-used shearing and converging motions, is much larger than the realistic values as observed at the photosphere. For example, \cite{Xing2024} used a driving speed with maximum value of around $20$~km~s$^{-1}$ which is larger than that in observations by over an order of magnitude.
This large boundary velocity weakly affects the quasi-static phase (where equilibrium is nearly achieved) and the eruption phase (where coronal velocities far exceed those at the lower boundary). However, the fast-driving speed may has a drastic effect during the slow-rise phase, since the driving speed is comparable to the typical slow-rise speed of a few tens~km~s$^{-1}$. Although \cite{2021NatAs...5.1126J} used a much smaller value of approximately $5$~km/s (which is smaller than typical slow-rise speed), it is still larger than actual photospheric values by a few times. As such, it remains to see how the magnitude of the driving speed affects the slow-rise phase.

To conduct MHD simulations using actual photospheric driving speed, however, is challenging in computation. Since the quasi-static energizing phase often takes hours to days~\citep{Zuccarello_2014}, which is much longer than the eruption phase, the entire evolution time $t_c$ (and the required computing time) is dominated by the pre-eruption evolution. The energy injection rate $P_s$ from the bottom surface is roughly proportional to the driving speed $v_d$, say $P_s = kv_d$ (where $k$ is a coefficient depending on specific magnetic configuration) and therefore, with a reduced driving speed, the time required for accumulation of the same amount of magnetic energy $E$ will increase proportionally, i.e., $t_c = E/P_s = E/(kv_d)$. Note that we have not yet considered the numerical diffusion (resistivity), which inevitably affects the accumulation of magnetic energy in the corona. The numerical diffusion depends on the nature of the numerical scheme (in both the implementation of the boundary boundary conditions and the solving the MHD equation in the corona) and the grid resolution. For a given scheme and grid resolution, the energy loss rate $P_n$ by numerical diffusion depends mainly on the magnetic field configuration (i.e., distribution of current density). For the sake of simplicity, let's further assume that energy flow across all other boundaries as well as magnetic energy conversion into other forms (e.g., thermal or kinetic energy) are negligible. Consequently, the accumulation of coronal magnetic energy is determined by the competition between energy injection from the bottom boundary and energy dissipation within the volume, i.e., the net increase rate of magnetic energy is $P=P_s-P_n$, and thus the simulation time should be
\begin{eqnarray}
\label{tc}
 t_c =\frac{E}{P}=\frac{E}{P_s-P_n}=\frac{E}{kv_d}\frac{1}{1-P_n/P_s}.
\end{eqnarray}
As can be seen, the time also depends on the ratio $P_n/P_s$. A too small driving speed $v_d$ may result in a surface energy injection rate $P_s$ comparable to the numerical diffusion rate $P_n$, and the time will significantly increase according to \Eq~\ref{tc}. Even worse, the simulation will fail when $P_s \leqslant P_n$, i.e., the energy injection rate is overtaken by the numerical diffusion, and no free magnetic energy can be accumulated in the corona. This is why many previous simulations chose to use an amplified driving speed to shorten the computing time and, more importantly, to ensure that $P_s \gg P_n$ to avoid the failure by the numerical diffusion. Clearly, if using a realistic driving speed, it requires the computation having a sufficient small numerical diffusion.

In this paper, based on the high-accuracy simulation of \citet{2021NatAs...5.1126J}, we managed to perform a series of experiments with the bottom driving speeds unprecedentedly approaching the photospheric values, which provides an unique opportunity to study how the slow-rise phase is influenced by the magnitude of the driving speed.

\section{Simulations}

\cite{2021NatAs...5.1126J} examined the initiation of solar eruptions through MHD simulations, demonstrating that slow shearing motion at the photosphere of a bipolar magnetic field can quasi-statically form an internal current sheet. The advantage of this simulation is that the numerical diffusion of the code, combined with sufficiently high resolution, is small enough to make magnetic energy loss negligible during the quasi-static energizing phase. With the current sheet becomes sufficiently thin, rapid magnetic reconnection occurs, triggering the eruption~\citep{2022A&A...658A.174B, jiangDevelopmentsFundamentalMechanism2024}.

Initially, the magnetic field is modeled as a potential arcade. Rotational flows were continuously applied to both polarities at the base, primarily at their edges, to inject free energy into the coronal field. This surface driving motion generates a magnetic configuration with substantial shear above the polarity inversion line (PIL). Such shearing methods are widely used in numerous 3D simulations to energize the coronal field before eruption~\citep[e.g.,][]{Amari2003,DeVore2008,Aulanier2010}.
This technique preserves the vertical component of the photospheric magnetic field distribution while enabling the self-consistent evolution of the horizontal components under the influence of applied shearing flows, as described by the induction equation. It ensures that the potential field energy remains constant while facilitating an increase in free energy as the magnetic field evolves.
 The MHD simulation solves the full set of 3D, time-dependent ideal MHD equations except that a small kinetic viscosity ($\nu = 0.05\Delta x^2/\Delta t$ where $\Delta x$ and $\Delta t$ are the spatial resolution and time step, respectively) is included in the momentum equation. The details of the MHD equations can be found in \citet{2021NatAs...5.1126J}. The initial temperature is uniform, with a typical coronal value of $T = 10^{6}$~K. The initial plasma density is uniform in the horizontal direction and vertically stratified by solar gravity. The magnetic flux distribution at the bottom boundary is characterized by a bipolar field composed of two Gaussian functions,
\begin{eqnarray}
B_z(x,y,0) = B_0e^{\frac{-x^{2}}{{\sigma}^{2}_x}}(e^\frac{-(y^{2}-{y}^{2}_c)}{{\sigma}^{2}_y}-e^\frac{-(y^{2}+{y}^{2}_c)}{{\sigma}^{2}_y}),
\end{eqnarray}
{\noindent}where $B_0 = 30$~G, $\sigma_x = 28.8 $~Mm, $\sigma_y = \sigma_x/4$, and $y_c = 11.5$~Mm, $\sigma_x$ and $\sigma_y$ determine the spread of the magnetic flux in the $x$ and $y$ directions, respectively, while $y_c$ dictates the separation between the two magnetic polarities along the $y$-axis.
 \begin{figure}
 	\centering
 	\includegraphics[width=0.5\textwidth]{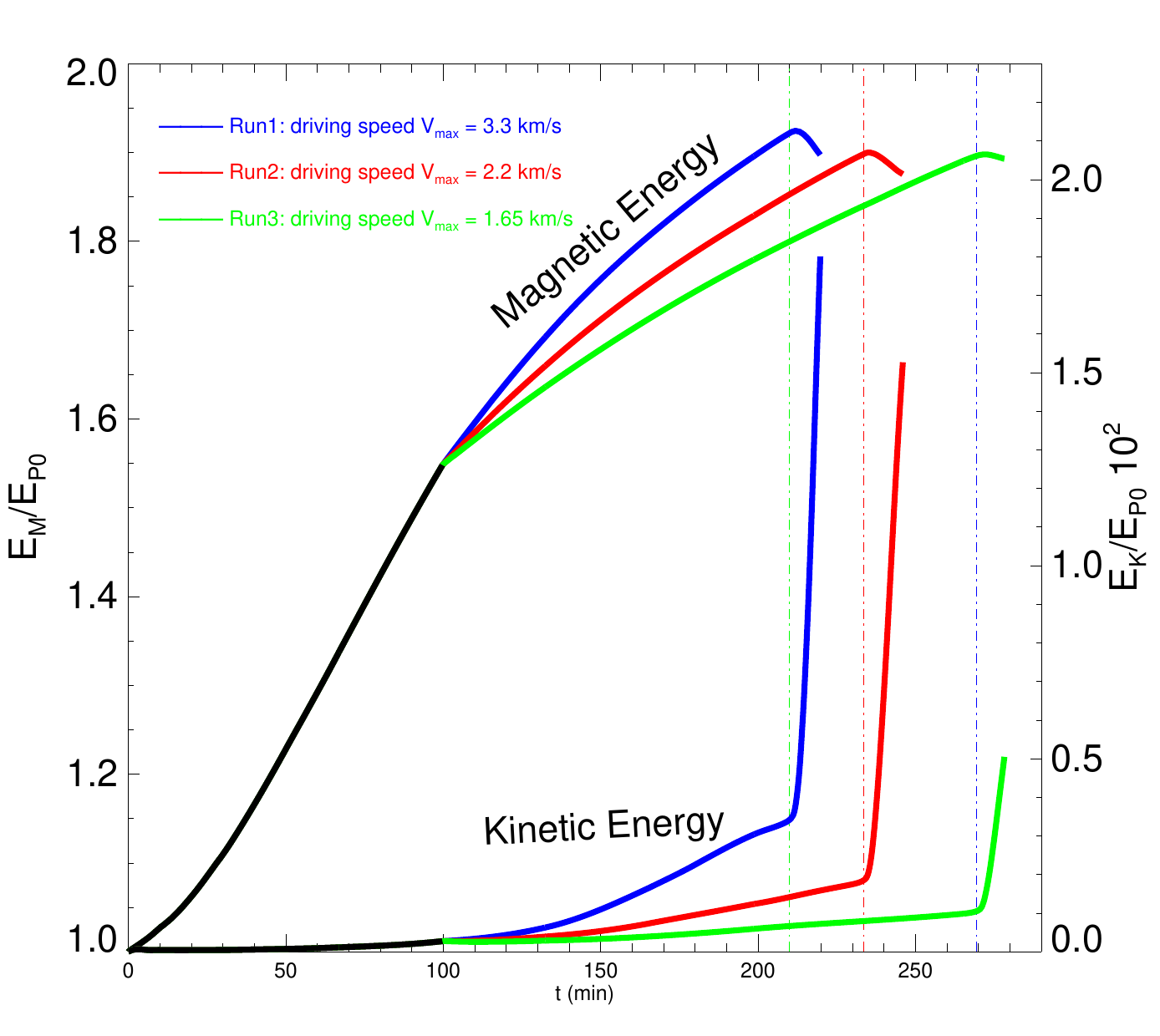}
 	\caption{Temporal evolution of magnetic and kinetic energies in the simulations. From $t=0$ to $100$~min, the simulation is driven by a driving speed with maximum of $5.5$~km~s$^{-1}$. Then the different simulations are shown with blue, red and green colors with maximum driving speed of $3.3$~km~s$^{-1}$, $2.2$~km~s$^{-1}$, and $1.65$~km~s$^{-1}$, respectively. The energies are normalized by the magnetic energy of the initial potential field $E_{P0}$. In all of the panels, the vertical dashed lines denote the onset time of eruptions, respectively. }
 	\label{fig1}
 \end{figure}

\begin{figure*}
	\centering
	\includegraphics[width=\textwidth]{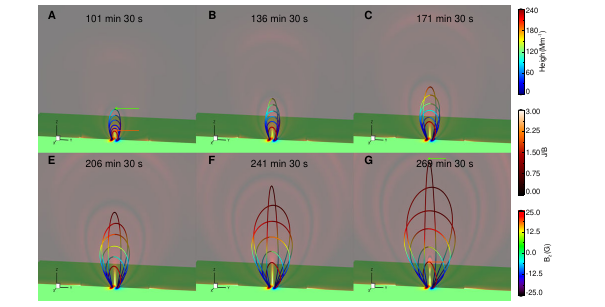}
	\caption{The evolution of the magnetic field lines in the simulation with  maximum of bottom driving speed as $1.65$~km~s$^{-1}$. The colored thick lines represent magnetic field lines. The background on the bottom and the vertical slice is plotted to show the vertical magnetic component $B_z$ and the ratio of current density to magnetic field strength $J/B$.}
	\label{fig2}
\end{figure*}

Our computational domain extends from $-360$~Mm to $360$~Mm in the transverse directions and from $0$ to $720$~Mm in the vertical direction, and is  resolved with an AMR grid with the highest resolution of $180$~km. At the bottom boundary ($z = 0$ plane), we impose slow-driving rotational flows at the field's footpoints. The velocity profile is defined by
\begin{eqnarray}
v_x=\dfrac{\partial\Psi(B_z)}{\partial{y}},v_y=-\dfrac{\partial\Psi(B_z)}{\partial{x}}
\end{eqnarray}
where
\begin{eqnarray}
\label{eq:psi}
\Psi(B_z)=v_0{B}^{2}_z e^\frac{-({B}^{2}_z-B^{2}_{z,\max})}{B^{2}_{z,\max}}
\end{eqnarray}
with $B_{z,\max}$ representing the maximum value of $B_z(x, y, 0)$ and $v_0$ a coefficient that controls the magnitude of the flow.
At the bottom boundary, the magnetic induction equation is directly solved to update the magnetic field self-consistently, driven by the surface flow. At the side and top surfaces: the tangential components of the magnetic field are linearly extrapolated from internal points, while the normal component is adjusted to satisfy the divergence-free condition, thereby minimizing numerical magnetic divergence near the boundaries.

We first apply the driving flow with a maximum speed of $5.5$~km~s$^{-1}$ until $t = 100$~min, which is well before the start of a slow-rise phase. Then, starting from $t = 100$~min, we carried out three simulations with the same settings except that the driving speed is reduced (by adjusting $v_0$ in \Eq~\ref{eq:psi}). The three simulations, referred to as Run1, Run2, and Run3, have maximum driving speed of $3.3$~km~s$^{-1}$, $2.2$~km~s$^{-1}$, and $1.65$~km~s$^{-1}$, respectively. By reducing the driving speed to approach the realistic photospheric values, we can examine how the magnitude of the bottom driving flow affects the evolution of the system, and more importantly, whether the appearance of the slow-rise phase before the eruption onset is related to the driving speed. We do not stop the driving flow throughout the simulation, since in observation the photospheric motion is ceaseless no matter there is eruption or not.

\begin{figure}
	\centering
	\includegraphics[width=0.5\textwidth]{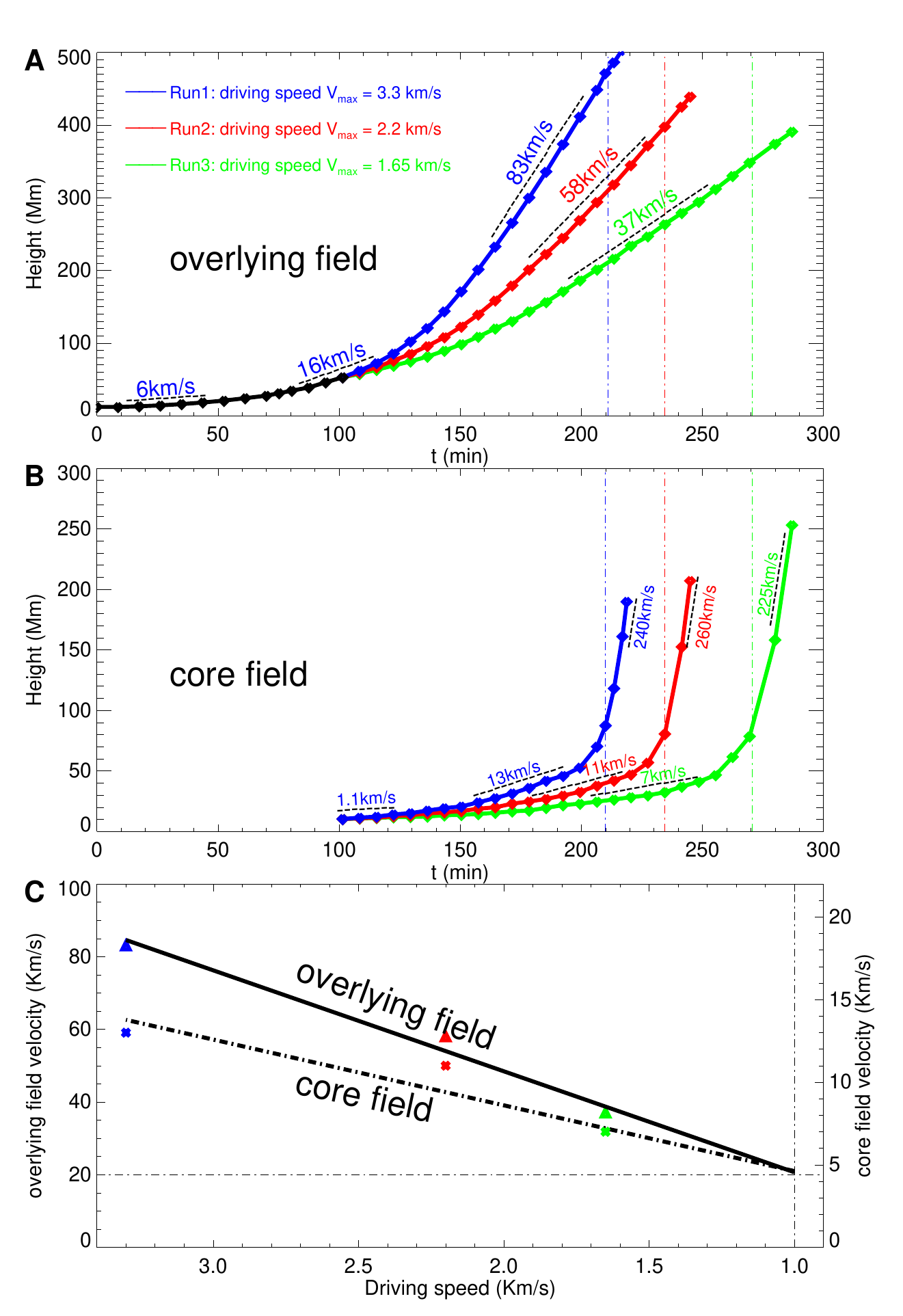}
	\caption{Height variation and the rising speed of two representative field lines, one for the overlying field and the other for the core field (which are denoted by the green and red arrows in \Fig~\ref{fig2}), in the different simulations. The blue, red, and green solid lines represent maximum driving speed of $3.3$~km~s$^{-1}$, $2.2$~km~s$^{-1}$, and $1.65$~km~s$^{-1}$, respectively. (A) The height of the overlying field line. (B) The height of the core field. The vertical dashed lines mark the times of eruption onset. The averaged velocity in the slow-rise phase is shown. (C) Variation of the slow-rise speed with the driving speed. The red, blue, and green triangles (crosses) respectively mark the slow-rise velocities of the overlying (core)field. The black lines illustrate the extrapolated velocities of overlying and core field lines in response to the diminishing bottom-driving speed.}
	\label{fig3}
\end{figure}

\section{Results}
\Fig~\ref{fig1} presents the magnetic and kinetic energy evolution profiles for the three simulations with different driving speeds (note that from $t=0$ to $100$~min, they have the same driving speed). The different simulations show the same behavior of free magnetic energy accumulation before eruption and release during eruption. The magnetic field evolves in accordance with what has been shown in \cite{2021NatAs...5.1126J}: before the eruption, the core field slowly expands outward with a current sheet gradually formed within; then, magnetic reconnection occurs at the current sheet, triggering the eruption and leading to a rapid release of magnetic energy.
In each simulation, the eruption onset time, as indicated by the dashed lines in \Fig~\ref{fig1}, is identified by the beginning of a sudden increase in the kinetic energy (which corresponds to the start of the magnetic energy decrease). Note that in the different simulations, the onset times of eruption are different. This is because right before the onset of eruption the accumulated magnetic energy should be approximately of the same amount while the energy injection rates (which depends on the driving speed) are different. As has been explained in \cite{2021NatAs...5.1126J}, to form a current sheet, the magnetic energy should be not far away from the open field energy. Indeed, in all the simulations, the magnetic energies right before the eruption onset are very close to each other (which is around $1.9 E_p$).

From the kinetic energy evolution, three phases can be identified. The first one is the quasi-static phase (from the beginning to around $t=100$~min), in which there is negligible change in the kinetic energy. After then and before the eruption onset time, the kinetic energy starts to increase noticeably but still gradually. This mild increase of kinetic energy is likely associated with the slow-rise phase in observations. The eruptive phase is marked by a sharp increase in kinetic energy. Since we focus on the slow-rise phase before eruption (and to save computing time), all the simulations are stopped shortly after eruption onset. Therefore, at the end of the simulations, the erupting field is still undergoing rapid accelerating and has not reached very high in the computational volume. Comparing the three simulations reveals that the kinetic energy during the slow-rise phase has different magnitudes, decreasing incrementally with smaller driving speeds, indicating that the driving speed affects the slow-rise phase.

\Fig~\ref{fig2} illustrates the evolution of a few magnetic field lines at different times in the Run3 simulation (with maximum driving speed of $1.65$~km~s$^{-1}$). These field lines are integrated accurately with their footpoints moving with the surface driving flow, thus they track the evolution of the same set of field lines with time. Before the slow-rise phase, the entire field expands slowly with roughly the same rate, but during the slow-rise phase, the field expands with rather different rates. Especially, when approaching the time when the current sheet is formed (\Fig~\ref{fig2}G), the expansion of the overlying field is significantly faster than that of the core field. Therefore to comprehensively analyze the slow-rise phase, we need to inspect the behavior of different field lines, in particular, differentiating the overlying field and the core field.

To quantify how the field expansion is influenced by the driving speed, in \Fig~\ref{fig3} we plot the height-time curves of the overlying and core fields in the three simulations by tracking two representative field lines. For the overlying field, a field line is traced from the center of the magnetic polarity (i.e., the point with the largest $B_z$) on the bottom surface. Since the driving velocity is zero at the polarity center, the footpoint of this overlying field line at different times is simply fixed. As this field line is not sheared, it can represent well the overlying field. This field line is shown by the highest one, (pseudo-colored by the height and marked by the green arrow) in each panel of \Fig~\ref{fig2}. The field line can be easily traced from the beginning of the simulation, i.e., $t=0$.
For the core field, we first locate a footpoint of the MFR axis at the early phase of the eruption (note that here the MFR forms during the eruption), and then we track the motion of this footpoint as driven by the surface driving flow at the bottom surface. Consequently a field line can be traced from this moving footpoint at different times. The evolution of this field line is shown by the lowest one (colored solely in red, and marked by the red arrow) in each panel of \Fig~\ref{fig2}. Since this field line undergoes significant shearing in the pre-eruption phase, we only traced it in time back to $t=100$~min.

As shown in \Fig~\ref{fig3}, the three phases can be clearly seen by the expansion of the two field lines. From the overlying field line, one can see the quasi-static and the slow-rise phases. Before $t \approx 100$~min, the field line expands quasi-statically with speed of around $6\sim 16$~km~s$^{-1}$, which is on the same order of magnitude of the maximum driving speed $5.5$~km~s$^{-1}$. Then after $t=100$~min, the field expansion is accelerated significantly, with speeds increased to approximately $80$~km~s$^{-1}$, $50$~km~s$^{-1}$, and $30$~km~s$^{-1}$, respectively, in the three simulations. These speeds are higher than the corresponding maximum driving speeds by an order of magnitude, and thus this phase can be taken as the slow-rise phase.
This suggests that the slow-rise phase is a generic phenomenon in the simulation, although it is modulated by the driving speed.~\footnote{It should be noted that when the driving speed becomes smaller, the slow-rise motion is also influenced more by the viscosity in the simulation, which reduces the slow-rise speed. }
From the core field line, one can see the three phases with distinct speeds. In the quasi-static phase, this core field line expands with around $1$~km~s$^{-1}$, somewhat slower than the driving speed. In the slow-rise phase it expands with speeds of $13$~km~s$^{-1}$, $11$~km~s$^{-1}$, and $7$~km~s$^{-1}$, a few times (around 5) of the corresponding driving speeds. So in both the two pre-eruption phases, the core field line expands much slower than the overlying one. These speeds are then rapidly accelerated to over $200$~km~s$^{-1}$, marking the onset of eruption. Since the eruption flow (and the core field) has not reached the heights of the overlying field line in the end of our simulations, the overlying field line is still in the slow-rising speed even passing the onset time of the eruption. After a few minutes when the core field catch up the overlying one from behind, the overlying field will also be accelerated rapidly as driven by the core one.

We further compare the slow-rise speeds of our simulation with an event: the slow-rise phase of an X1.0 flare eruption that occurred on 2021 October 28 in AR 12887. This event has been carefully analyzed by~\citet{duanInitiationMechanismFirst2023} and also~\citet{10.1093/mnras/stae2088}. They measured the slow-rise speeds of the overlying loops and filament (corresponding to the core field) and eliminated the projection effect by using a triangulation approach based on the SDO and STEREO observations as well as coronal field extrapolation analysis. It is found that the actual slow-rise speed of  the overlying loops and the filament are around 5~km~s$^{-1}$ and 20~km~s$^{-1}$, respectively. As shown in \Fig~\ref{fig3}C, a simple extrapolation of the results of the three simulations suggests that if with driving speed of $1$~km~s$^{-1}$ (which is typically the realistic value in the photosphere), the slow-rise speeds of the overlying field and the core field would also be around 20~km~s$^{-1}$ and 5~km~s$^{-1}$, respectively, closely matching the observed values.


\section{CONCLUSION}

In the context of \citet{2021NatAs...5.1126J}'s simulations, which have established a fundamental mechanism for solar eruptions, we have lifted the model to a higher level of reality by reducing the speed of bottom-driving flow to closely approaching the actual photospheric value. Our simulations show that the mechanism of eruption initiation is not sensitive to the driving speed (that is, all the different simulations produce the eruption in the same way), while the slow-rise phase is indeed influenced by the magnitude of the driving speed. By incrementally reducing the bottom-driving speed, we observed a correspondingly slower increase in the kinetic energy growth curve during the slow-rise phase. We further traced the expansion of the magnetic field lines which clearly exhibit slow rising before the eruption onset. The overlying field line above the core flux has a slow-rise speed always larger than the driving speed by over an order of magnitude, and it is modulated by the driving speed. The core field also expands with speed much higher than the driving speed but much lower than that of the overlying field. By incrementally reducing the bottom-driving speed to realistic photospheric values, we anticipate better matches between simulated slow-rise speeds and some observed ones. Finally, it is worth noting that our findings are only applicable to the cases where the pre-eruptive structure is a sheared arcade and reconnection does not occur before eruption. It remains to be investigated in future study whether the findings are also correct for cases where the pre-eruptive structure is a flux rope, or where there is a pre-eruptive magnetic reconnection.



\normalem
\begin{acknowledgements}
This work is jointly supported by National Natural Science Foundation of China (NSFC 42174200), Shenzhen Science and Technology Program (Grant No. RCJC20210609104422048), Shenzhen Key Laboratory Launching Project (No. ZDSYS20210702140800001), Guangdong Basic and Applied Basic Research Foundation (2023B1515040021). We thank the reviewer for helping to improve the manuscript significantly.
\end{acknowledgements}


\end{document}